\documentclass[sigconf]{acmart}

\usepackage{booktabs} % For formal tables
\usepackage{balance}
\usepackage{amssymb}
\usepackage{amsmath}
\usepackage[]{algorithm}
\usepackage[noend]{algpseudocode}
\setcopyright{rightsretained}
\begin{document}

\copyrightyear{2017}
\acmYear{2017}
\setcopyright{acmcopyright}
\acmConference{CIKM'17 }{November 6--10, 2017}{Singapore, Singapore}\acmPrice{15.00}\acmDOI{10.1145/3132847.3132860}
\acmISBN{978-1-4503-4918-5/17/11}

\title{Data Driven Chiller Plant Energy Optimization with Domain Knowledge}

%\author{
%Kok Soon Chai$^{\dag}$ \and Bryan Keating$^{\ddag}$ \and Hoang Dung Vu$^{\dag}$ \and Marianne Winslett$^{\ddag}$ \and Kaige Yang$^{\flat}$ \and Xiaoyan Yang$^{\P}$ \and Zhenjie Zhang$^{\P}$ \\
%\and $^{\dag}$Kaer Pte. Ltd. \\
%\{koksoon.chai,jose.vu\}kaer.com
%\and $^{\ddag}$University of Illinois at Urbana Champaign\\
%\{bkeatin2,winslett\}@illinois.edu
%\and $^{\flat}$University of College London\\
%zczlkya@ucl.ac.uk
%\and $^{\P}$Advanced Digital Sciences Center\\
%\{xiaoyan.yang,zhenjie\}@adsc.com.sg
%}

\author{Hoang Dung Vu}
\affiliation{%
  \institution{Kaer Pte. Ltd.}
}
\email{jose.vu@kaer.com}

\author{Kok Soon Chai}
\affiliation{%
  \institution{Kaer Pte. Ltd.}
}
\email{koksoon.chai@kaer.com}

\author{Bryan Keating}
\authornote{This work was done during the author's internship with Advanced Digital Sciences Center.}
\affiliation{\institution{University of Illinois at Urbana Champaign}}
\email{bkeatin2@illinois.edu}

%\author{Marianne Winslett}
%\affiliation{\institution{University of Illinois at Urbana Champaign}}
%\email{winslett@illinois.edu}

\author{Nurislam Tursynbek}
\authornote{This work was done during the author's internship with Advanced Digital Sciences Center.}
\affiliation{\institution{Nazarbayev University}}
\email{nurislam.tursynbek@nu.edu.kz}

\author{Boyan Xu}
\authornote{This work was done during the author's internship with Advanced Digital Sciences Center.}
\affiliation{\institution{Guangdong University of Technology}}
\email{xuboyan@mail2.gdut.edu.cn}

\author{Kaige Yang}
\authornote{This work was done during the author's internship with Advanced Digital Sciences Center.}
\affiliation{\institution{University College London}}
\email{zczlkya@ucl.ac.uk}

\author{Xiaoyan Yang}
\affiliation{\institution{Advanced Digital Sciences Center}}
\email{xiaoyan.yang@adsc.com.sg}

\author{Zhenjie Zhang}
\affiliation{\institution{Advanced Digital Sciences Center}}
\email{zhenjie@adsc.com.sg}

\renewcommand{\shortauthors}{H. D. Vu et al.}

%\authornote{The secretary disavows any knowledge of this author's actions.}
%\affiliation{%
%  \institution{Institute for Clarity in Documentation}
%  \streetaddress{P.O. Box 1212}
%\email{webmaster@marysville-ohio.com}
%
%\author{Lars Th{\o}rv{\"a}ld}
%\authornote{This author is the
%  one who did all the really hard work.}
%\affiliation{%
%  \institution{The Th{\o}rv{\"a}ld Group}
%  \streetaddress{1 Th{\o}rv{\"a}ld Circle}
%\email{larst@affiliation.org}
%
%\author{Lawrence P. Leipuner}
%\affiliation{
%  \institution{Brookhaven Laboratories}
%  \streetaddress{P.O. Box 5000}}
%\email{lleipuner@researchlabs.org}
%
%\author{Sean Fogarty}
%\affiliation{%
%  \institution{NASA Ames Research Center}
%  \city{Moffett Field}
%  \state{California}
%  \postcode{94035}}
%\email{fogartys@amesres.org}
%\affiliation{%
%  \institution{Palmer Research Laboratories}
%  \streetaddress{8600 Datapoint Drive}
%\affiliation{\institution{The Th{\o}rv{\"a}ld Group}}
%\email{jsmith@affiliation.org}
%
%\author{Julius P.~Kumquat}
%\affiliation{\institution{The Kumquat Consortium}}
%\email{jpkumquat@consortium.net}
%
%% The default list of authors is too long for headers}

\begin{abstract}
Refrigeration and chiller optimization is an important and well studied topic in mechanical engineering, mostly taking advantage of physical models, designed on top of over-simplified assumptions, over the equipments. Conventional optimization techniques using physical models make decisions of online parameter tuning, based on very limited information of hardware specifications and external conditions, e.g., outdoor weather. In recent years, new generation of sensors is becoming essential part of new chiller plants, for the first time allowing the system administrators to continuously monitor the running status of all equipments in a timely and accurate way. The explosive growth of data flowing to databases, driven by the increasing analytical power by machine learning and data mining, unveils new possibilities of data-driven approaches for real-time chiller plant optimization. This paper presents our research and industrial experience on the adoption of data models and optimizations on chiller plant and discusses the lessons learnt from our practice on real world plants. Instead of employing complex machine learning models, we emphasize the incorporation of appropriate domain knowledge into data analysis tools, which turns out to be the key performance improver over state-of-the-art deep learning techniques by a significant margin. Our empirical evaluation on a real world chiller plant achieves savings by more than 7\% on daily power consumption.
\end{abstract}

%\begin{CCSXML}
%<ccs2012>
%  <concept_id>10010520.10010553.10010562</concept_id>
%  <concept_desc>Computer systems organization~Embedded systems</concept_desc>
%  <concept_significance>500</concept_significance>
%  <concept_id>10010520.10010575.10010755</concept_id>
%  <concept_desc>Computer systems organization~Redundancy</concept_desc>
%  <concept_significance>300</concept_significance>
%  <concept_id>10010520.10010553.10010554</concept_id>
%  <concept_desc>Computer systems organization~Robotics</concept_desc>
%  <concept_significance>100</concept_significance>
%  <concept_id>10003033.10003083.10003095</concept_id>
%  <concept_desc>Networks~Network reliability</concept_desc>
%  <concept_significance>100</concept_significance>
%</ccs2012>
%\end{CCSXML}
%
%\ccsdesc[500]{Computer systems organization~Embedded systems}

% We no longer use \terms command

%\keywords{ACM proceedings, \LaTeX, text tagging}

\maketitle

\section{Introduction}\label{sec:intro}

Energy management is known as one of the most crucial problems in the vision of smart building  \cite{manic2016building}. On globe scale, 30\% of total energy consumption and 60\% of electricity consumption are spent on buildings \cite{manic2016building}. In order to reduce the carbon emission and enhance environment sustainability, it is highly demanding to redesign and deploy the new generation of building management systems (BMS), which aims to minimize the building energy consumption by controlling and scheduling the components in the buildings based on accurate models on energy behavior.

Among the components controlled by BMS in modern commercial buildings, chiller plant is recognized as the most energy hungry component. In Singapore, statistics reveal that about one third of the building energy is spent by air-conditioning equipments in households \cite{SES}. Despite of the long history of chiller plant optimization techniques, it remains a challenging problem to generate the desirable cooling load for demands from the building while minimizing the total energy consumption. The challenges mainly come from two difficulties, including the difficulty on modelling a non-stable dynamic physical system, as well as the difficulty on accurate measurement of the running status of the chiller plant. Conventional optimization techniques developed in mechanical engineering community mostly rely on physical models of the equipments, based on over-simplified assumptions on the running conditions. Real world chiller plants are hardly compatible with these physical models, due to the variance coming from the difference on installation and ageing of equipments.
%There is no engineer in real world following such methodology when operating and optimizing chiller plants.}

\begin{figure}
	\centering
	\includegraphics[width=.8\columnwidth]{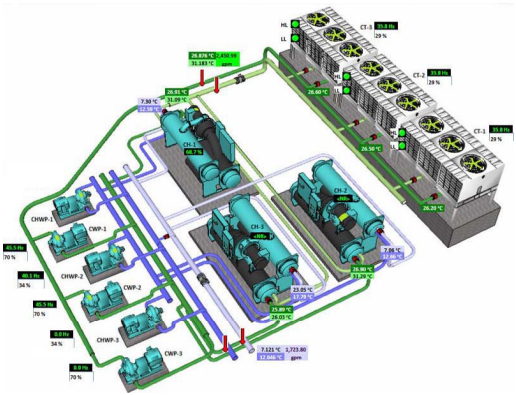}\\
	\vspace{-5pt}
	\caption{Interface of Kaer's chiller plant monitoring system \emph{K-realtime}.}\label{fig:chiller}
	\vspace{-10pt}
\end{figure}

Until recently, thanks to the quick advances of sensor technologies, for the first time, chiller plant system can accurately monitor all of its components in real time, as is shown in Figure \ref{fig:chiller}. The explosive growth of data collected from chiller plant, as well as the increasing analytical power, are now raising new interests on data-driven approaches on chiller plant optimization. By utilizing the historical and fresh data from the sensors, data analysis is expected to enhance the performance of the existing optimization techniques.

One key advantage of data-driven optimization approach is its high adaptivity to varying external conditions, ageing devices and human behaviors, all of which are known as important performance factors but hard to model using physical models only. To fully exploit such advantage, on the other hand, we must resort to powerful data analytical tools to fully capture these variables not covered by traditional physical models. Blind optimization over the massive data records, may not improve over dull rule-based strategies commonly used in existing building management systems (BMS) \cite{doukas2007intelligent}. Instead, the success of data-driven approaches heavily relies on the appropriate integration of domain knowledge from experts and the valuable data insight unknown to domain experts. DeepMind \cite{Deepmind}, for example, is attempting to optimize the power for cooling service in Google's data centers, based on the patterns of energy consumption linked to running status of the machines in the data center.

%Deep learning, especially Recurrent Neural Network and its variants, e.g., \cite{hochreiter1997long}, is recognized as the most promising technique applicable to time series domains, such as handwriting \cite{graves2009novel}, speech audio \cite{sak2014long} and traffic pattern \cite{liang2016mercury}. Different from rule-based strategy commonly used in existing building management systems (BMS) \cite{doukas2007intelligent}, deep learning builds models on top of historical observations over the physical system, without any prior knowledge on the underlying mechanism. It can easily integrate additional information from outside of the target system. DeepMind \cite{Deepmind}, for example, is attempting to optimize the cooling service in Google's data centers, based on the running status of the machines in the data center.

In this paper, our discussions on data-driven chiller plant optimization are in three-fold. First, we present how we solve the chiller plant energy optimization problem following a carefully chosen technical roadmap, including three steps, i.e., active data enrichment, energy consumption prediction and real-time configuration search. Each subproblem solved in the steps of the roadmap is designed to fully exploit corresponding data available to the system, thus rendering more accurate models over the targets, such as future cooling load and chiller energy consumption. Second,  we vertically decompose the whole chiller plant architecture into a number of modules, such that accurate prediction output of a module could be used as inputs to other modules. Each module is then independently analyzed by training models using the sensor data. Given the high accuracy of the data models over individual modules and the low dependency between the modules, thanks again to our domain expertise, the combination of the models generates excellent overall performance on every subproblem in our technical roadmap. Third, we discuss how our domain knowledge is used to guide our model selection. Instead of using complex deep learning models, e.g., Recurrent Neural Network \cite{hochreiter1997long}, our system adopts certain simple yet useful models based on our understanding to the general running mechanisms of the equipments, avoiding the under-fitting problems with over-complicated deep learning models.

%We show that stacked neural network helps to improve the accuracy of all prediction tasks, without using much background knowledge on the specifications of the equipments in chiller plant. Such accuracy improvement is due to the time-relevant dependencies between the running status of the equipments in the system, captured by recurrent neural network models. Secondly, we introduce how we build our own complete solution to machine-learning based chiller plant optimization system, including the offline data archive module, model re-training module and online real-time decision making module. We show that machine learning itself is insufficient to tackle the whole problem. It is equivalently important to carefully design the interface between the modules and interpret the results in the right manner. By implementing the model over the system architecture, we empirically evaluate our solution on real chiller plants, including analysis on performance gains and discussions on potential performance limitations.

To elaborate the details of our technique, the rest of the paper is organized as following. Section \ref{sec:related} reviews existing studies on chiller plant optimization and deep learning over time series domain. Section \ref{sec:datap} discusses the data acquisition and pre-processing. Section \ref{sec:model} presents the models employed to predict cooling load and energy consumption. Section \ref{sec:opt} introduces the real-time optimization system we implemented to apply our optimization algorithms. Section \ref{sec:exp} empirically evaluate the usefulness of our models and system, and Section \ref{sec:conclu} finally concludes the paper.

\section{Related Work}\label{sec:related}

In this section, we review the existing studies on a variety of relevant research problems, including the models on the cooling loads and energy performance, control approaches using neural network and extreme seeking, as well as data-driven optimization techniques for other Internet of Things (IoT) systems.

Chiller plant optimization is a traditional research topic in mechanical engineering. A large number of research works in the literature attempt to model the behavior of chiller plant based on the physical rules of the cooling devices, e.g., \cite{zhao2012review,yang2005line,iwafune2014short,ben2004cooling,li2009applying,li2014building,momtazpour2015analyzing}. Most of these works, however, do not consider the varying factors, such as ageing equipments and indoor activities.

Neural networks are considered as an option for modeling in mechanical engineering community \cite{xu2005optimization,wang2008supervisory,chow2002global}. These works directly include neural network in their prediction models. Such approach may not reflect the actual running mechanism beneath the equipments. Moreover, due to the existing control policy of chiller plant, the data collected by the sensor network usually only spans on a subspace of configuration space. Training over such data may cause high generalization errors. In our work, we solve these problems with active data enrichment and select models based on domain knowledge over the equipments.

Performance measurement is an equally important problem to optimization in chiller plant, in the sense that a systematic approach is needed to evaluate the improvement of new optimization strategies adopted by the chiller plant. This is not a simple task, because the external variables keep changing over time. Reduced energy consumption could be fully due to better weather but not good control optimization. It is therefore demanding to build a general baselining technique, which enables to estimate the energy consumption of a chiller plant if an old control strategy is employed, such as the methods proposed in\cite{jelali2006overview,salsbury2015two}. Our technique proposed in this paper can also be used as a baselining approach. In our empirical evaluations, we use our well trained model for energy baselining.

Extreme seeking is an active and dynamic search approach, designed to seek for optimal system configuration even when the system is constantly in unstable state, e.g., \cite{tyagi2006extremum,mu2016optimization}.  However, existing results show that extreme seeking may not always find the optimum in the search space, because of the dynamic nature of the chiller plant system. We handle the problem simply by hard coding domain knowledge into the optimization procedure. Such simple strategy turns out to be highly effective on finding nearly optimal control decisions.

%\cite{wei2014modeling}.

Recent years witness the quick advances of deep learning techniques. Recurrent neural network (RNN) \cite{sak2014long,graves2009novel,hochreiter1997long,liang2016mercury} and its variants are the most popular models for time series domain. However, RNN is well known for its high complexity and low training efficiency. An under-fitted RNN model, i.e., without sufficient training, fails to fully exploit the power of the deep neural networks. In this paper, we solve the problem by decomposing the chiller plant system into small modules and choose appropriate simple models for individual modules based on our domain knowledge. We show that the accuracy performance is excellent and the training efficiency of the approach is much better than that of complex neural networks.

Internet of Things (IoT) technique is quickly growing, thanks to the cheap and reliable data acquisition methods enabled by new generation of sensor networks. Predictive maintenance, for example, is one of the most important applications of IoT. In \cite{jung2017vibration}, Jung et al. show that analytics over vibration sensors on motors and tubes can accurately predict the problems with the equipments. Such analysis supports timely replacement of the equipments, instead of traditional strategies with fixed replacement period. In our work, as is shown in the experiments, with accurate model on the running status of the equipments, our analysis also supports real-time fault diagnosis.

%\section{Preliminary}\label{sec:prelim}

\section{Preliminaries}\label{sec:datap}
%\subsection{Chiller Plant}

\begin{figure}
	\centering
	\includegraphics[width=.78\columnwidth]{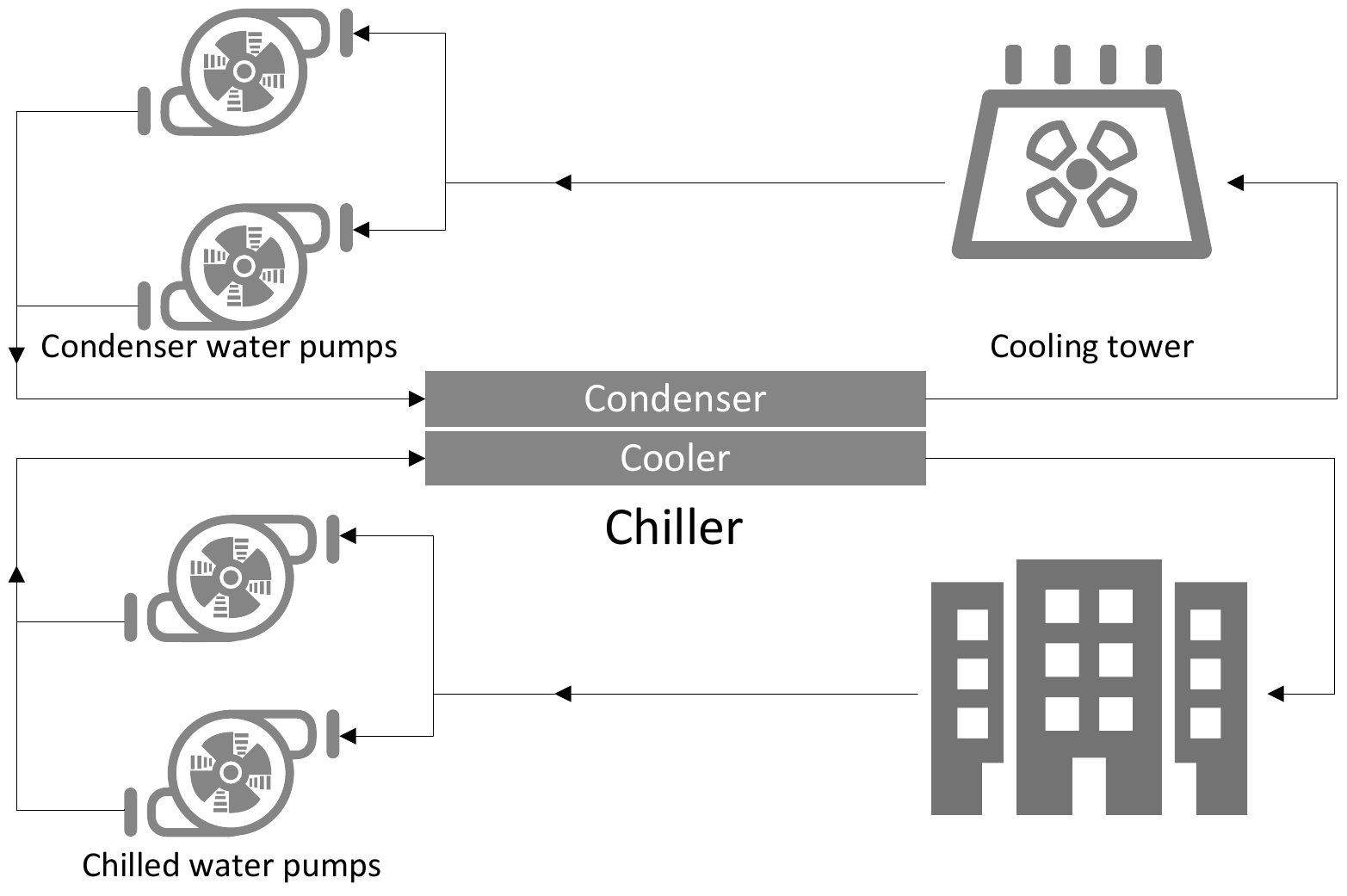}\\
	\vspace{-5pt}
	\caption{Abstract structure of chiller plant.}\label{fig:sys}
	\vspace{-10pt}
\end{figure}

Chilled water-based cooling systems are commonly used to cool and dehumidify air in various types of large buildings, such as offices, industry sites, hospitals, schools and etc. \cite{chillerdesign}. A typical chiller plant (chilled water-based) is equipped with four types of energy consuming equipments, including chillers, cooling towers, condenser water pumps and chilled water pumps. In Figure \ref{fig:chiller}, the example chiller plant consists of \emph{three} chillers, \emph{three} cooling towers, \emph{three} condenser water pumps and \emph{three} chilled water pumps. To better understand how chiller plant works, we present a simplified structure of chiller plant in Figure \ref{fig:sys}. There are two independent cycles in every chiller plant. The inner cycle pushes the chilled water from chillers to the indoor space of the building for air cooling. The chilled water absorbs the heats from the building and runs back to the chiller at higher temperature. Similarly, the outer cycle pushes the condenser water from chiller to the cooling tower, which emits the heat to the outdoor environment, with colder water coming back to the chiller. The heat exchange between the inner and outer cycles is executed by the chillers. The chilled water and condenser water flow in the cycles, powered by the chiller water pump and condenser water pump respectively. To simplify the presentations in the rest of the paper, we summarize the notations in Table \ref{tb:notation} following most of the research works on chiller plant optimization.

\begin{table}
	\caption{Table of notations}\label{tb:notation}
	%\small
	\centering
	\begin{tabular}{|l|p{2.5in}|}
		\hline
		\textbf{Symbol} & \textbf{Description} \\ \hline\hline
		CH & Chiller \\
		CT & Cooling tower \\
		CWP & Condenser water pump \\
		CHWP & Chilled water pump \\
		CWFM &  Condenser water flow model \\
		CHFM &  Chilled water flow model \\
		CWTM &  Condenser water temperature model \\
		VSD & Variable speed drive \\
		$cwp\_speed$ &  VSD speed of condenser water pump \\
		$ct\_speed$ &   VSD speed of cooling tower fan \\
		$chwp\_speed$ & VSD speed of chilled water pump \\
		kW & Kilowatt \\
		$chkw$ &    Chiller power in kW \\
		$ctkw$ &    Cooling tower power in kW \\
		$cwpkw$ &   Condenser water pump power in kW \\
		$chwpkw$ &  Chilled water pump power in kW \\
		RT & Refrigeration ton \\
		RLA & Rated load amperage \\
		$chfhdr$ & chilled water flow in/out chillers \\
		$cwfhdr$ & condenser water flow in/out chillers \\
		$cwshdr$ & condenser water temperature into chillers \\
		$chsp$ & chiller setpoint indicating the desired chiller water temperature \\
		$db$ & dry bulb temperature \\
		$rh$ & relative humidity \\
		\hline
	\end{tabular}
\end{table}

%input feature category

\subsection{Data Collection}\label{sec:datacollect}

Generally speaking, chiller plant control system collects three types of data, including control data, external condition data and sensor data.

\noindent\textbf{Control Data:}In a chiller plant, equipments like cooling tower and water pumps are controlled by variable-speed drive (VSD)~\footnote{https://en.wikipedia.org/wiki/Adjustable-speed\_drive} , which is an inverter to control the speed of motors. Using Kaer's K-realtime system, we record the change of the running parameters, as listed below:

\begin{itemize}
	\item VSD speed of condenser water pump in percentage
	\item VSD speed of cooling tower fan in percentage
	\item VSD speed of chilled water pump in percentage
	\item Binary Configurations (On/Off) of chillers
	\item Binary Configurations (On/Off) of condenser water pumps
	\item Binary Configurations (On/Off) of chilled water pumps
	\item Binary Configurations (On/Off) of cooling towers
\end{itemize}

\noindent\textbf{External Condition Data: }Besides control parameters, weather data are also collected including the relative humidity and the dry bulb temperature. One measurement is collected in every minute, in order to track immediate change of weather conditions.
%We also record the power consumption of each component, cooling load in refrigeration tons (RTs) and heat balance of the system. All sensor readings are collected every second.

%In total, a dataset of $6K$ samples (each sample contains the above list of sensor readings) from January 2016 to June 2016 (?) is collected for study. Table~\ref{tb:notation} summarizes the notations used in the rest of the paper and their physical meanings.

\noindent\textbf{Sensor Data:  }Sensors are deployed on all equipments in chiller plants. Specifically, we record the power of each equipment, i.e., chillers, pumps and cooling towers, the input and output temperatures of the chilled water and condensed water, the flow rate of the running water. We also record the power, cooling load in refrigeration tons (RTs) and heat balance of the system. All sensor readings are collected every minute. The details of sensor data are described in Table \ref{tb:notation}.

\subsection{Technical Target and Roadmap}

Based on high school physics, the cooling load generation of the chiller plant, known as refrigeration ton (or RT in short), is proportional to the flow rate of the chilled water as well as the temperature difference between output and input chilled water. Therefore, in order to generate the desired cooling load, the major tradeoff is on the chilled water flow rate and the temperate of the input chilled water into the building. The flow rate of the chilled water is mainly controlled by the chilled water pump, which increases the flow rate by consuming more electricity. The temperate of the input chilled water is decided by the power consumption of the chiller, as well as the condenser water pump and the cooling tower. Basically, the system could spend more energy either on speeding up chilled water pump, or lowering the temperate of the chilled water. The key of energy saving is to find the balance between these two factors, i.e., a configuration achieving the cooling load and minimizing total power consumption over all equipments.

In this paper, we mainly target to optimize the chiller plant energy consumption by applying \emph{micro-control} strategies over the equipments. As discussed in Section~\ref{sec:datacollect}, there are two types of control parameters in a chiller plant: VSD speed of water pumps and cooling tower fans, and configurations (On/Off) of equipments. We do not consider the on/off decisions made over the equipments, i.e., chiller, pumps and cooling towers. This is because, in practice, most of the \emph{macro-control} policies over the number of simultaneous running equipments are based on long-term cooling load estimations. Building managers easily make optimal macro-control decisions, by pre-scheduling the configurations of equipments on a daily basis. Based on our testing results, machine learning and data analysis do not outperform humans on equipment scheduling, mainly due to the limited scheduling options. In Figure \ref{fig:chiller}, for example, there are only three chillers available to the plant, thus generating only seven possible running chiller combinations.

\begin{figure}
	\centering
	\includegraphics[width=.7\columnwidth]{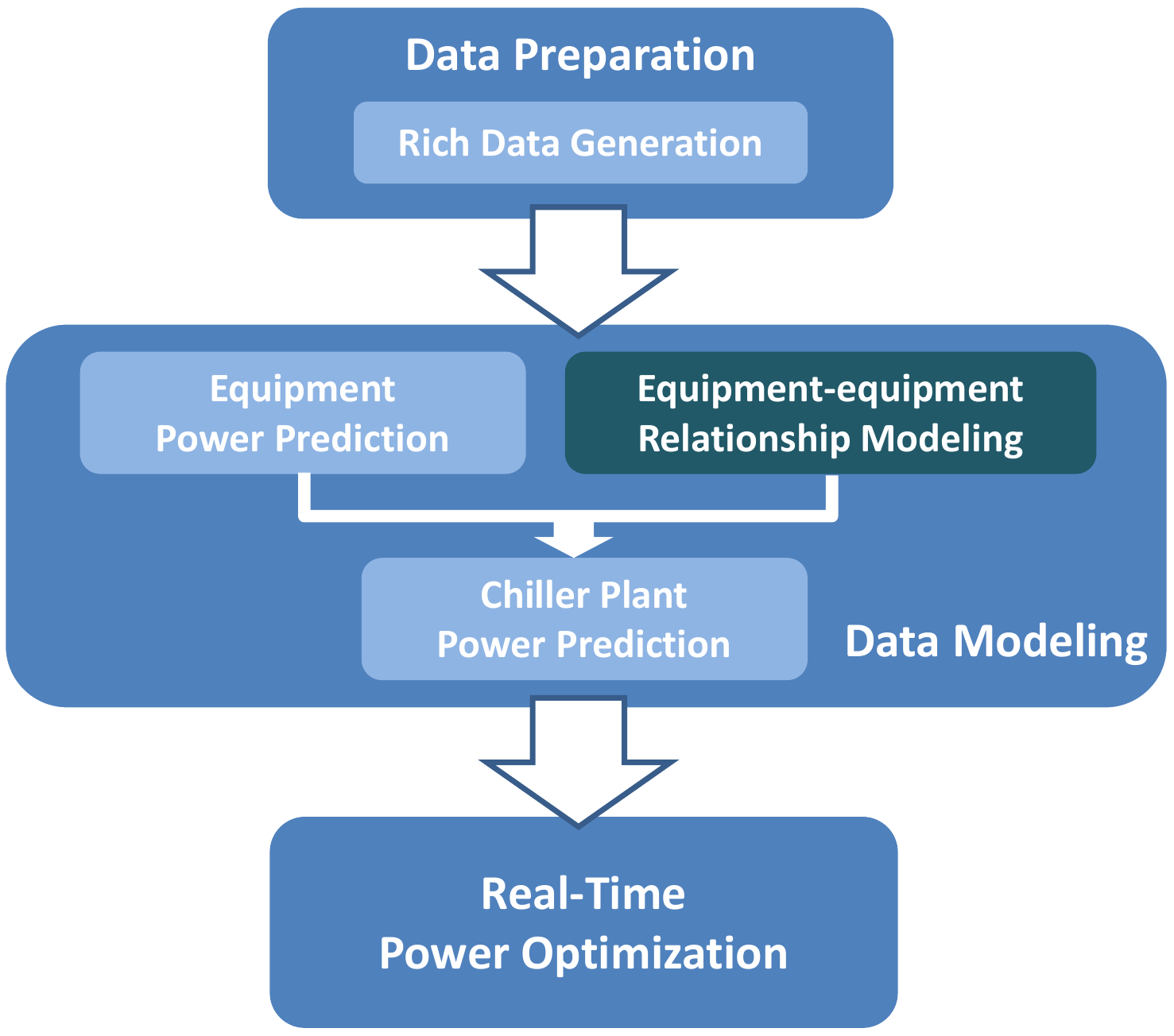}\\
	\vspace{-5pt}
	\caption{Technical roadmap of our data-driven optimization approach.}\label{fig:roadmap}
	\vspace{-10pt}
\end{figure}

Based on our observation above, we focus on the optimization over micro-control strategies, by fine tuning of the controlling parameters, i.e., VSD of the equipments. In order to fully optimize micro-control, we follow the technical roadmap as is shown in Figure \ref{fig:roadmap}. There are three major technical components in our approach. In \emph{data proprocessing}, we attempt to enrich the data to overcome the difficulty of generalization. In \emph{data modeling}, we decompose the chiller plants into modules and build data models over the modules in an independent manner. There are two types of modules, used to capture power and relationships of equipments respectively, for decomposition purpose. In \emph{real-time optimization}, we deploy the power model on a controller computer directly connected to the chiller plant, which makes decisions on parameters on the fly. The technical details of these components are available in the following sections.

\section{Data Techniques}\label{sec:model}

%In this section, we discuss our techniques used to handle data, particularly on data preprocessing and data modelling.

\subsection{Data Enrichment in Preprocessing}\label{sec:model:enrich}
%\subsubsection{Problematic Data - Lacking of Richness}

%\begin{figure}
%  \centering
%  % Requires \usepackage{graphicx}
%  \includegraphics[width=2.7in]{fig/poordatafit}\\
%  \caption{Fitting on 慞oor?dataset versus fitting on 慠ich?dataset: The device (which one?) is originally operated in the shaded speed region (from 20\% to 35\%), which results in a wrong model (the red dots).}\label{fig:poordata}
%\end{figure}

\begin{figure}[ht]
	\centering
	%\begin{tabular}{c}
	\includegraphics[width=.73\columnwidth]{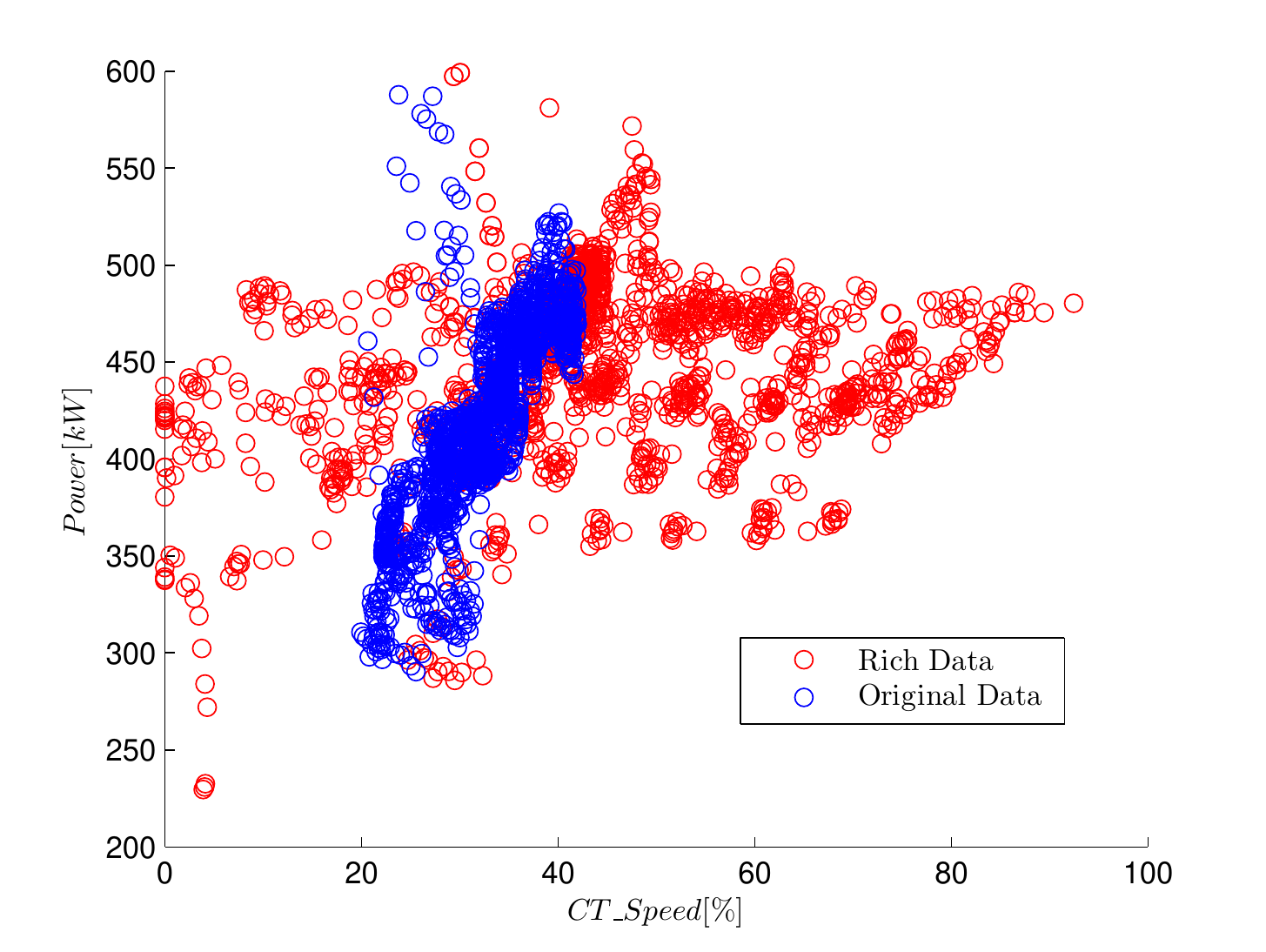}  %\\
	%   \includegraphics[width=.7\columnwidth]{fig/cwp_d}
	%\end{tabular}
	\vspace{-5pt}
	\caption{The original data over power and cooling tower speed spans a clear linear subspace, because of the fixed VSD in the configuration. The model built on top of the data may not generalize well beyond the subspace. By enriching the data, the new model over rich is capable of estimating the power consumption when other VSD is applied.}\label{fig:richdata}
	\vspace{-5pt}
\end{figure}

The lack of generality in the data is an important problem, which could be easily overlooked.
%which does not attract attentions in almost all existing studies on chiller plants.
Simple data modelling over the existing chiller plant data may result in useless model with high generalization error. In an extreme case, a chiller plant always runs at a fixed configuration, e.g., fixed VSD speed for pumps and fans. By training over data from such chiller plant, the resulting data model is only applicable to the current or similar configurations, and does not generate meaningful prediction for other varying configurations.
%This leads to largely redundant data, making it difficult for machine learning methods to capture the variations or dynamics of the system.
Figure~\ref{fig:richdata} plots the data distribution over the cooling tower speed $CT\_Speed$ and the total system power, collected in fully controlled chiller plant in 2016 with a fixed VSD configuration setting (denoted as original data) and random VSD configuration (denoted as rich data) respectively. In the original data, the cooling tower fan is mainly operated at the speed between $20\%$ to $40\%$ of the maximum speed. With fixed VSD configuration settings for all devices, the total system power and cooling tower speed clearly span on a linear subspace. %Although $CWP\_Speed$ covers almost the full range, its relationship with $syskw$ (blue dots) covers a subspace compared with that of the rich data (red dots) (Figure~\ref{fig:richdata}(b)). Aside from the noise in $syskw$, figure~\ref{fig:richdata} shows that $CT\_Speed$, $CWP\_Speed$, and $syskw$ form a degenerate subspace.
The results show that data model using fixed VSD configuration does not have much generalization capability when other configurations are used by the chiller plant.

%An example of poor dataset that gives wrong models is shown in Fig.1. In this scenario, the device is always operated in the shaded region (from 20\% to 35\%). If these data (the 憄oor?data in the figure) is used for fitting, it will result a wrong model (the red dots). It is very important for machine learning to have a reasonable 慻ood?training dataset.

%To tackle this problem, we first review the richness of historical chiller plant data. Particularly, we examine if the data covers the full ranges of control parameters. If not, we proceed to excite the chiller plant to generate rich data.

%\subsubsection{Generating Rich Data}

To tackle the problem, we randomly update each control parameter of the chiller plant so as to explore the full range of values. As shown in Figure~\ref{fig:richdata}, the rich data spans a much larger space, providing more accurate insights into the behavior of the equipment and offering better opportunity for modeling and optimization.

%we need some data enrichment scheme. Two methods are proposed to generate rich data: \emph{Signal Sweeping} and \emph{Random}.

%\noindent\textbf{Signal Sweeping} gradually sweeps each control parameter from its minimum value to its maximum value as shown in Figure~\ref{fig:datagen}(a). However, it gets more time-consuming as the number of parameters increases.
%
%\noindent\textbf{Random walk} uses random walk to let each control parameter explore its full range of values as shown in Figure~\ref{fig:datagen}(b). This method is less time-consuming than signal sweeping. But it may not cover all the data range.

%\begin{figure}
%\centering
%\hspace{-15pt}
%\begin{tabular}{cc}
%  % after \\: \hline or \cline{col1-col2} \cline{col3-col4} ...
%  % Requires \usepackage{graphicx}
%  \includegraphics[width=.45\columnwidth]{fig/gen_rich_data1} &   \includegraphics[width=.45\columnwidth]{fig/gen_rich_data2} \\
%  (a) signal sweeping & (b) random walk  \\
%\end{tabular}
%  \vspace{-5pt}
%  \caption{Data enrichment schemes.}\label{fig:datagen}
%  \vspace{-5pt}
%\end{figure}

\subsection{Overall Model over Chiller Plant}

\begin{figure}[t]
	\centering
	\includegraphics[width=1.1\columnwidth]{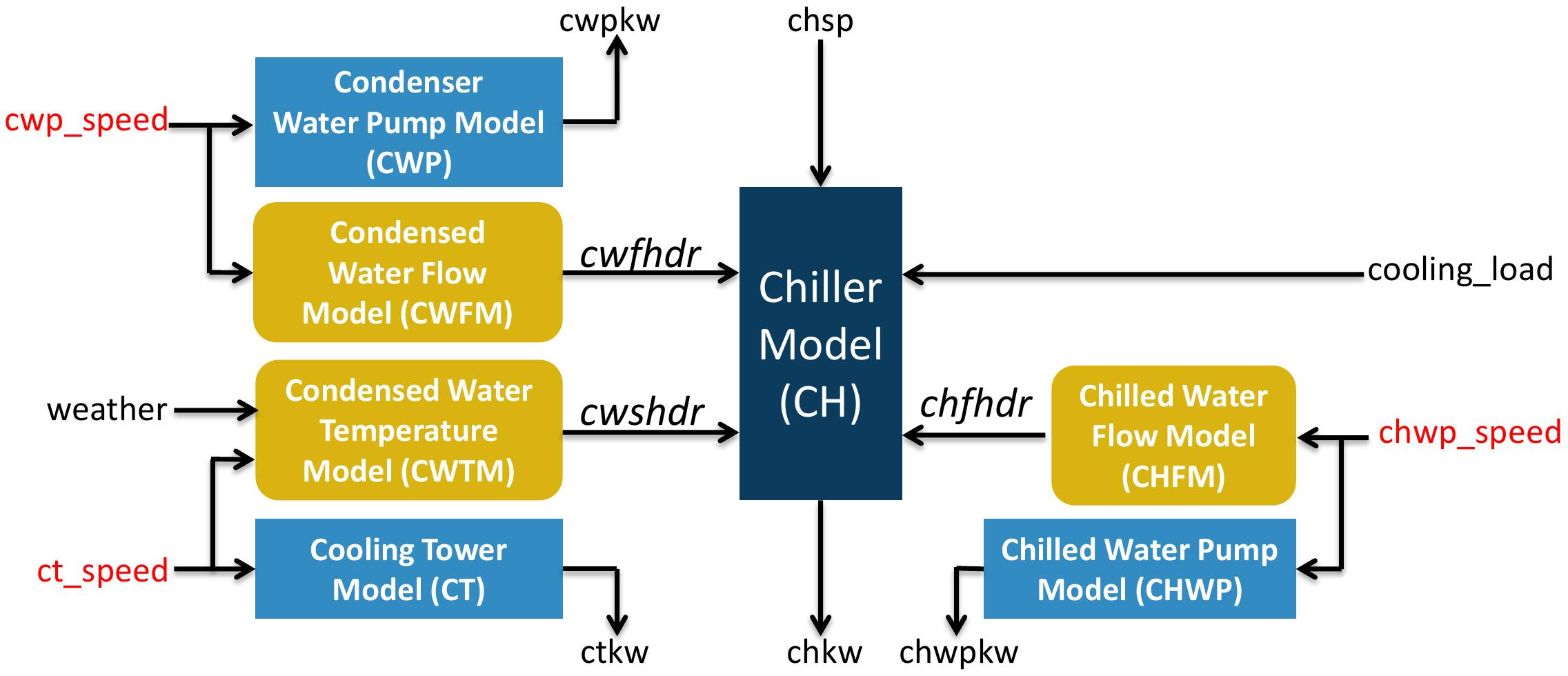}\\
	\vspace{-5pt}
	\caption{Module-wise modeling of chiller plant.}\label{fig:framework}
	\vspace{-5pt}
\end{figure}

Figure~\ref{fig:framework} depicts the overall idea of chiller plant decomposition for data modelling. A chiller plant is decomposed into modules, each of which corresponds to a block in Figure~\ref{fig:framework}. Each block also represents a module-wise data model with the incoming edges as the input variables and the outcoming edges as output/prediction variable. The connections among the modules, i.e., the prediction results fed from one module to others, are designed based on our understanding to the mechanism beneath chiller plant. Therefore, the overall model reflect our domain knowledge over the equipments.

Specifically, there are two types of modules in the overall model. Rectangle blocks denote energy consumption models over single equipments. %Round cornered blocks denote behavior models predicting particular properties of the equipments.
Round cornered blocks denote relationship models predicting particular properties of the system that affect two types of equipments.
For instance, \emph{Condenser Water Flow Model} (CWFM) is used to predict the flow rate of the condenser water running through the condenser water pumps and chillers. Chiller is singled out in the figure, because it is the most complex equipment in chiller plant system. Particularly, the model of the chillers utilizes the outputs from others, including the flow rate of the condenser water and the chilled water, and the temperature of the condenser water flowing into the chillers. Notice that, as shown in Figure~\ref{fig:sys}, chillers are affected by condenser water and chilled water flowing in and out. However, temperature of the condenser water flowing out of the chillers, and temperatures of the chilled water flowing in/out of the chillers are not modeled (Figure~\ref{fig:framework}). This is because they are captured in the cooling load of the system, which is also an input to the chiller model. In this work, we assume the current cooling load of the chiller plant is known and changes slowly over time.

In the rest of the section, we abuse CWP, CHWP, CT, CH, CHFM, CWFM and CWTM to denote the seven modules (Figure~\ref{fig:framework}) respectively. In next subsection, we discuss how to pick up right \emph{data models} for individual modules.

\subsection{Models over Modules}
The modules in a chiller plant model, as is shown in Figure \ref{fig:framework}, are treated in two different ways, according to the number of inputs: single input single output (SISO) and multiple input single output (MISO). Chillers, for example, adopts MISO model. Generally, chiller is the core and most complex equipment in a chiller plant, whose performance is affected by all other modules in the plant. All SISO modules follow Affinity Laws \cite{affinitylaw}. Based on this observation, we apply different models to the modules to match the complexity of the modules. In Table~\ref{tb:model}, we summarize the models employed in our system over individual modules. The rest of the subsection elaborates the models in detail.

\begin{table}
	\small
	\caption{Summary of models over modules.}\label{tb:model}
	\centering
	\begin{tabular}{|l|l|l|l|}
		\hline
		Module & Type & Model Type & Prediction Variable \\ \hline\hline
		CWP & SISO & Polynomial & Power \\
		CHWP & SISO & Polynomial & Power \\
		CT & SISO & Polynomial & Power \\ \hline
		CH & MISO & MLP & Power \\
		CHFM & MISO & MLP & Water flow \\
		CWFM & MISO & MLP & Water flow \\
		CWTM & MISO & MLP & Water temperature \\
		\hline
	\end{tabular}
\end{table}

\noindent\textbf{Affinity Laws}
The Affinity Laws of centrifugal pumps or fans describes the relationship between power $P$ and shaft speed $S$ with impeller diameter held constant \cite{affinitylaw}:
%\begin{enumerate}
%\item Flow is proportional to shaft speed: $$\frac{Q_1}{Q_2} = \left(\frac{N_1}{N_2}\right)$$
%\item Pressure or Head is proportional to the square of shaft speed:
%$$\frac{H_1}{H_2} = \left(\frac{N_1}{N_2}\right)^2$$
%\item Power is proportional to the cube of shaft speed:
%$$\frac{P_1}{P_2} = \left(\frac{N_1}{N_2}\right)^3$$
%\end{enumerate}
\begin{itemize}
	\item Power is proportional to the cube of shaft speed:
	$$\frac{P_1}{P_2} = \left(\frac{S_1}{S_2}\right)^3$$
\end{itemize}
The affinity laws ensures that it is sufficient to use polynomial regression to model power of water pumps and cooling tower fans in a chiller plant. There is no need to apply more complex models.

\noindent\textbf{Polynomial Regression for SISO Modules}
According to affinity laws of pumps and fans, we apply polynomial regression to model the heads and power of water pumps and cooling tower fans, the equations of which can be given as follows:
\begin{equation}
y = \alpha_0 + \alpha_1x + \alpha_2x^2 + ... + \alpha_kx^3 + \epsilon
\end{equation}
where, %$n \in \{2, 3\}$ according to affinity laws, and

\begin{tabular}{ccl}
	%\hline
	$y$ & : & predicted value of power \\
	$x$ & : & control parameter  \\
	$a_i$ & : & regression coefficient, $i\in \{0, 1, ..., k\}$ \\
	$\epsilon$ & : & error term
	%\hline
\end{tabular}

\noindent The input and output of SISO models CT, CWP and CHWP are summarized as follows:

\vspace{0.05in}
\qquad \qquad
\begin{tabular}{|l|l|l|}
	\hline
	Model & $x$ & $y$ \\ \hline
	CT & $ct\_speed$ & $ctkw$ \\
	CWP & $cwp\_speed$ & $cwpkw$ \\
	CHWP & $chwp\_speed$ & $chwpkw$ \\
	\hline
\end{tabular}
\vspace{0.05in}
%\noindent Theoretically, the relationship between power and shaft speed follows third degree polynomial regression according to affinity laws, we find quadratic regression sufficient to model the data of SISO components, i.e., $k=2$. Least squares is applied to estimate the regression coefficients.

\begin{figure}
	\centering
	\includegraphics[width=2in]{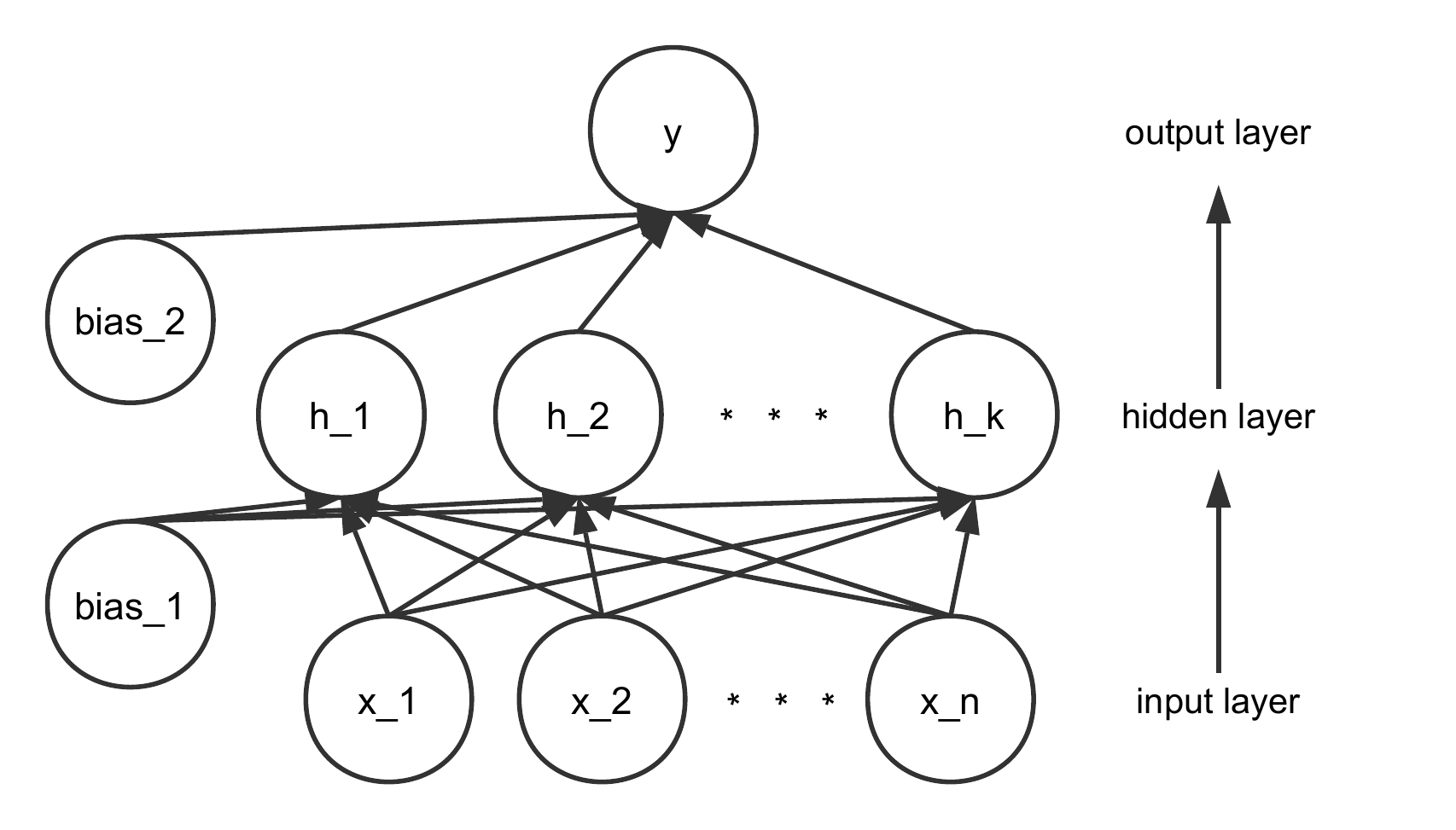}\\
	\caption{One hidden layer MLP}\label{fig:mlp}
\end{figure}

\noindent\textbf{MLP for MIMO Modules} A multilayer perceptron (MLP) is a feedforward artificial neural network (ANN) that learns a function to map input features $\{\textbf{x}\}$ to a target output $y$. Figure~\ref{fig:mlp} shows a one hidden layer MLP. We apply MLP to model the relationships between VSD speed and water flow rate and temperature.
%For each module, we deploy a one hidden layer MLP with three hidden nodes and logistic activation function.
The input and output of each module is defined as follows.

\vspace{0.05in}
%\qquad \qquad
\begin{tabular}{|l|l|l|}
	\hline
	Model & $\textbf{x}$ & $y$ \\ \hline
	CHFM & $chwp\_speed$, on/off of CHWPs, & $chfhdr$ \\
	CWFM & $cwp\_speed$, on/off of CWPs,  & $cwfhdr$ \\
	CWTM & $ct\_speed$, on/off of CTs, $db$, $rh$ & $cwshdr$ \\
	\hline
\end{tabular}
\vspace{0.05in}

\noindent \textbf{CHFM} is to predict the flow rate $chfhdr$ of chilled water flowing in/out the chillers. The input features $\textbf{x}_{CHFM}$ are VSD speed of chilled water pumps $chwp\_speed$ and configurations of chilled water pumps, i.e., on/off of chilled water pumps.

\noindent \textbf{CWFM} is to predict the flow rate $cwfhdr$ of condenser water flowing in/out the chillers. The input features $\textbf{x}_{CWFM}$ are VSD speed of condenser water pumps $cwp\_speed$ and configurations of condenser water pumps, i.e., on/off of condenser water pumps.

\noindent \textbf{CWTM} is to predict the condenser water temperature $cwshdr$ fed into the chillers. The input features $\textbf{x}_{CWTM}$ are weather (dry bulb temperature $db$ and relative humidity $rh$), VSD speed of cooling tower fans $ct\_speed$ and configurations of cooling towers. Notice that $cwshdr$ does not depends on condenser water pump speed.

\noindent\textbf{CH} Similarly, we apply MLP to model power of chillers. The output of CHFM, CWFM and CWTM are fed into CH models for chiller power prediction. The input features $\textbf{x}_{CH}$ of a chiller model CH are chilled water flow rate $chfhdr$, condenser water flow rate $cwfhdr$, temperature of condenser water fed into chillers $cwshdr$, the cooling load of the system and the setpoint $chsp$. The output of CH is chiller power $chkw$.
%The limited-memory Broyden-Fletcher-Goldfarb-Shanno (L-BFGS)~\cite{lbfgs} algorithm is applied to estimate the parameters.
%We deploy a one hidden layer MLP with three hidden nodes and logistic activation function to model the chillers.

\section{Real-time Power Optimization}\label{sec:opt}
The ultimate goal of this work is to minimize the total power consumption of a chiller plant, so as to minimize the operating cost, while still meeting the demanding cooling load. With all the models of the chiller plant, the power optimization problem can be formulated as follows:
%\begin{flalign}
%\underset{\textbf{x}}{\text{minimize}} \ \
%& J(\textbf{x}) = \sum_i CH_i(\textbf{x}_{CH}) + \sum_i CHWP_i(\textbf{x}_{CHWP}) \nonumber  && \nonumber \\
%& \qquad \ \ \ \ + \sum_i CWP_i(\textbf{x}_{CWP}) + \sum_i CT_i(\textbf{x}_{CT})
%  \nonumber
%\end{flalign}
\begin{flalign}
\underset{\textbf{x}}{\text{minimize}} \ \
& J(\textbf{x}) = \sum_i CH_i(\textbf{x}) + \sum_i CHWP_i(\textbf{x}) \nonumber  && \nonumber \\
& \qquad \ \ \ \ + \sum_i CWP_i(\textbf{x}) + \sum_i CT_i(\textbf{x})
\nonumber
\end{flalign}
\begin{flalign}
\text{subject to}\qquad & \underline{x}\leq x \leq \overline{x}, \quad \forall x \in \textbf{x} && \nonumber \\
& \underline{cwfhdr} \leq \hat{cwfhdr} \leq \overline{cwfhdr} \nonumber \\
& \underline{chfhdr} \leq \hat{chfhdr} \leq \overline{chfhdr} \nonumber \\
& \underline{cwshdr} \leq \hat{cwshdr} \leq \overline{cwshdr} \nonumber
\end{flalign}
%\begin{flalign}
%\text{subject to}\qquad & 0\leq s(\tau)\leq s_{\textup{max}}, \quad \forall \tau && \nonumber
%\\
%& \sum_{\tau=t}^{t+T-1} \gamma \beta(\tau)\leq N_{\textup{max}}
%\end{flalign}
where $\textbf{x} =\{cwp\_speed, chwp\_speed, ct\_speed\}$ is the set of control parameters .
%$\textbf{x}_B$ represents the set of input feature of block B.
$J(\textbf{x})$ is the predicted total power of the chiller plant. The constraints on controllable $x$ are to make sure that the cooling load of each chiller does not exceed its maximum cooling load, and the constraints on predicted water flow and temperature $\hat{cwfhdr}$, $\hat{chfhdr}$, $\hat{cwshdr}$, which are factors of chillers, are to prevent chillers from unnecessary performance fluctuation\footnote{These constraints require domain knowledge, and are therefore usually provided by the manufacturer of chillers.}.
Given weather conditions and the current state (e.g., cooling load, configurations of equipments\footnote{The configurations of equipments are followed pre-defined schedules. }) of the chiller plant, the optimizer tries to find the optimal control parameters $ct\_speed$, $cwp\_speed$, $chwp\_speed$ that minimize the total power while satisfy all constraints.
%In this study, we do not consider configurations (on/off) of chillers, water pumps and cooling towers as adjustable control parameters in optimization. The configurations of components are followed pre-defined schedules. 
A derivative-free optimization method, constrained optimization by linear approximation (COBYLA)~\cite{powell2007view}, is used to solve the optimization problem. In the real-time optimization applied on chiller plants, control parameters are updated every 2-3 minutes. 

Notice here meeting demanding cooling load is not explicitly modeled as constraints in the optimization problem. The reasons are twofold. First, the optimizer usually update the control parameters by a small amount in real-time that have little effect on the cooling load. Second, when a chiller plant is designed, designers will add (quite) some margin to the maximum cooling load obtained from energy audit. This leaves a large space for optimization. Therefore, it rarely fails to meet the cooling load during optimization, especially when the configurations of chillers are not changed.

\section{Empirical Evaluations}\label{sec:exp}
\subsection{Settings}

%1. add baselines to settings,

%2. combine tables together, three column for each method

%3. remove sub subsection, use description

%4. random data gen description

%5. optimization description

%\subsubsection{Data}
\noindent\textbf{Data} We evaluate our proposed power prediction models on real world data collected from the chiller plant shown in Figure~\ref{fig:chiller}, which supports cooling load service for a multi-building campus in Singapore. The data set consists of 12,520 samples of 15 days from January 24, 2017 to February 16, 2017. The sensor data are recorded in every minute. We divide the dataset into five folds, each consisting of data from three consecutive days, for five-fold cross validation.

%We use the latest 20\% for testing and the rest 80\% for training.

%It is randomly divided into two sets: 80\% for training and 20\% for testing.

During our data collection, data enrichment scheme (Sec \ref{sec:model:enrich}) is employed to enhance the generality of the data. To enrich data with minimal effect on normal chilling service, random update of control parameters is applied only a few times a day, lasting for half an hour each time. When the chiller plant is under testing, or some equipment is under repair or replacement, the data enrichment scheme is run for longer period of time, e.g., half a day.
%(how random method is applied here? - how long and how often)

%To generate rich data, we apply signal sweeping/random method on the chill plant for how long (?) to obtain a rich dataset of how many(?) records.

Because chiller plant is not a stable system, outliers commonly occur in the data set, especially on readings from the sensors. In data preprocessing, we apply random sample consensus (RANSAC)~\cite{ransac} to filter out the outliers.

\noindent\textbf{Metric}
The mean absolute percentage error (MAPE) is used to evaluate the performance of proposed prediction models. MAPE is formally defined as follows:
\begin{equation}
\text{MAPE} = \frac{100}{n} \sum_i^n \left|\frac{y_i - \hat{y}_i}{y_i}\right|,
\end{equation}
where $y_i$ is the actual value and $\hat{y}_i$ is the prediction outcome.

%\subsubsection{Baselines}
\noindent\textbf{Baselines} We use \textbf{DDO} to denote our own data-driven optimization approach, which uses polynomial regression (PR) for SISO modules and MLP for MISO modules. A one hidden layer MLP with 3 hidden nodes and logistic activation function is deployed. We compare DDO mainly with long short term memory (LSTM)~\cite{hochreiter1997long}, which is the state-of-the-art model for time series data prediction. LSTM is implemented in Keras 1.2.2 with Theano 0.8.2. The batch size is set as 128. We use MAPE as the loss function and run 2,000 epochs for each model training. For SISO modules and chillers, we use one-layer LSTM with 8 hidden state nodes. For other MISO modules, a three-layer LSTM is trained with the $1^{st}$/$2^{nd}$/$3^{rd}$ hidden layer having hidden state nodes that are $3/2/1$ times of the input size. The hyperparameter settings of LSTM are tuned for best performance.

\noindent\textbf{Hardware \& Software}
All our algorithms are implemented in Python, and applied to the chiller plant using an IPC (industrial PC) running on Intel Core i5-6599TE with 8GB of RAM, using Ubuntu 14.04 LTS. The communication with the chiller plant is through a Building Automation and Control networks (BACnet).

%\subsubsection{Metric}

\begin{figure}
	\centering
	\includegraphics[width=.6\columnwidth]{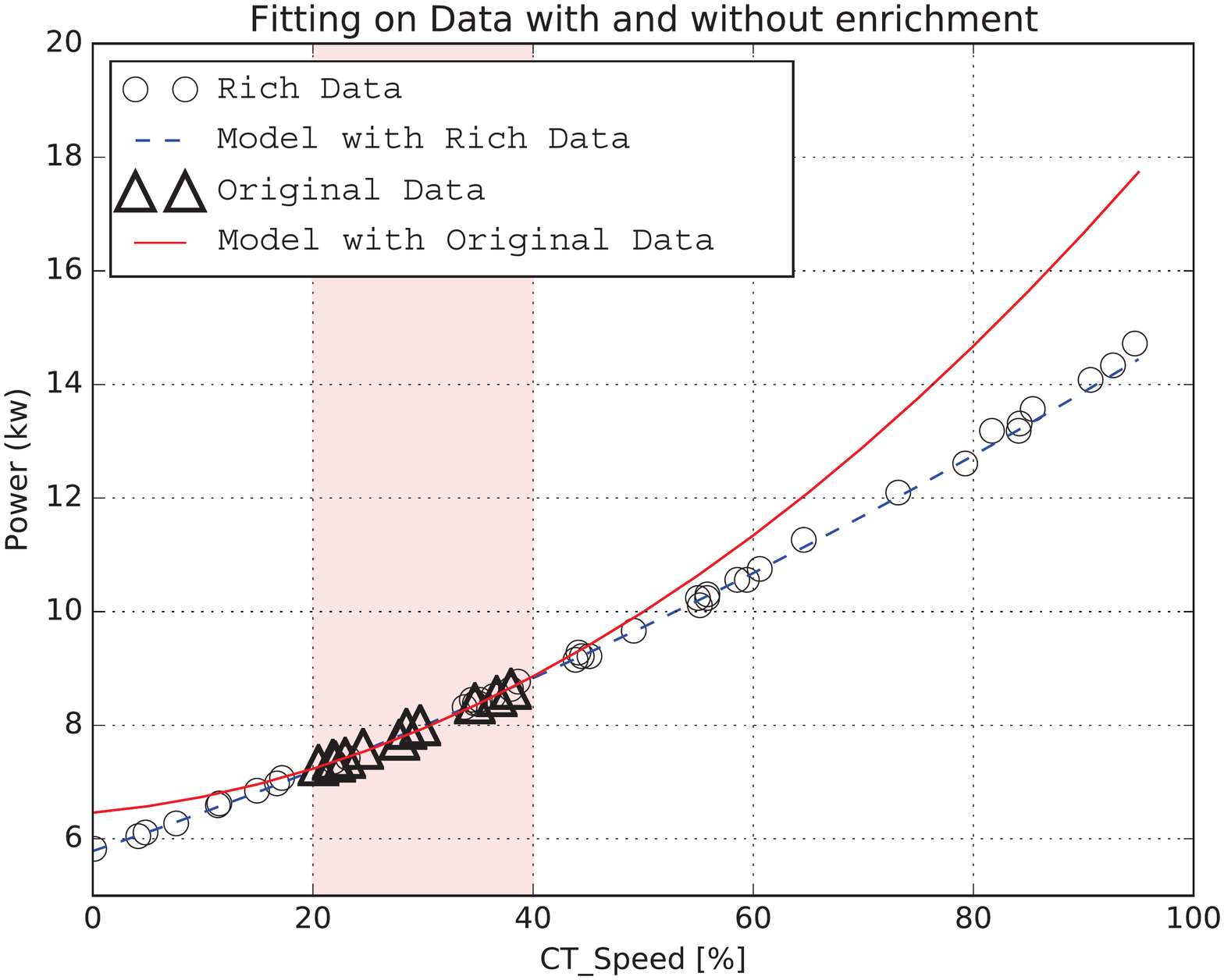}\\
	\vspace{-5pt}
	\caption{Comparison of Models }\label{fig:fig5}
	\vspace{-10pt}
\end{figure}

\subsection{Evaluation of Data Enrichment Scheme}

%\subsubsection{Effect of Rich Data on Prediction Accuracy}
We evaluate the effect of data enrichment on power prediction accuracy. Two PR models are trained using the data with and without enrichment in Figure~\ref{fig:richdata} for predicting the power of cooling tower. The results are shown in Figure~\ref{fig:fig5}. The model trained using the original data deviates significantly from the true data points in VSD ranges outside the original data, and therefore has a much higher MAPE ($7.25\%$) than that ($0.65\%$) of the model trained using rich data.
%achieves MAPE$=0.65\%$. However, the MAPE of model trained using data without enrichment is $7.25\%$, which is much higher.
This indicates data enrichment is extremely necessary before applying any chiller plant modeling.

\subsection{Evaluation of Module-Wise Models}

\begin{table}
	\small
	\caption{MAPEs(\%) of SISO models}\label{tb:mapesiso}
	\centering
	\scalebox{0.88}{
		\begin{tabular}{|l|c|c|c|}
			\hline
			Model & PR (DDO) & MLP  & LSTM \\ \hline\hline
			CHWP1 & 1.67 &  1.59 & 5.32 \\
			CHWP2 & 2.10 &  2.16 & 6.27 \\
			CHWP3 & 2.33 &  2.41 & 4.64 \\
			AVG\_CHWP  & \textbf{2.03} &     2.05 & 5.41 \\ \hline
			
			CWP1 &  1.48 &  1.43 &  2.16 \\
			CWP2 &  1.33 &  1.16 &  2.92 \\
			CWP3 &  1.02 &  1.02 &  1.79 \\
			AVG\_CWP  & 1.28 & \textbf{1.20} & 2.29 \\ \hline
			
			CT1 &   1.49 &  0.77 &  1.86 \\
			CT2 &   0.94 &  0.94 &  2.23 \\
			CT3 &   0.77 &  0.77 &  1.59 \\
			AVG\_CT  & 1.07 & \textbf{0.83} & 1.89 \\ \hline
			
			%    CHWP1 & 0.62  & 0.44 \\
			%    CHWP3 & 0.80 & 1.07 \\
			%    CHWP & \textbf{0.71}   & 0.76 \\ \hline
			%    CWP1 & 1.86   & 1.83 \\
			%    CWP3 & 0.80   & 0.49 \\
			%    CWP & 1.33    & \textbf{1.16} \\ \hline
			%    CT1 & 1.60   & 0.89 \\
			%    CT2 & 0.80  & 1.92 \\
			%    CT3 & 0.62  & 1.43 \\
			%    CT & \textbf{1.14}  & 1.41 \\
			%   \hline
	\end{tabular}}
\end{table}

%\begin{figure*}
%\begin{tabular}{ccc}
%  % after \\: \hline or \cline{col1-col2} \cline{col3-col4} ...
%  \includegraphics[width=2in]{fig/cwp1}
%  & \includegraphics[width=2in]{fig/anncwp1}
%  & \includegraphics[width=2in]{fig/cwp3} \\
%  (a) DDO (PR) & (b) MLP & (c) LSTM \\
%
%%  \includegraphics[width=2in]{fig/chwp1}
%%  & \includegraphics[width=2in]{fig/chwp2}
%%  & \includegraphics[width=2in]{fig/chwp3} \\
%%  (d) & (e) & (f) \\
%%
%%  \includegraphics[width=2in]{fig/ct1}
%%  & \includegraphics[width=2in]{fig/ct2}
%%  & \includegraphics[width=2in]{fig/ct3} \\
%%  (g) & (h) & (i) \\
%\end{tabular}
%\caption{Visualization of SISO CWP1 models on the testing data}
%\label{fig:siso}
%\end{figure*}

%\begin{figure*}
%  \centering
%  % Requires \usepackage{graphicx}
%  \includegraphics[width=7in]{fig/opt1}\\
%  \caption{Result of real-time optimization}\label{fig:optresult}
%\end{figure*}

\noindent\textbf{Evaluation of SISO Models} The power prediction results of SISO models -- PR, MLP and LSTM are summarized in Table~\ref{tb:mapesiso}. For each module, we report the MAPE of each equipment (e.g., CT1/2/3) and the average MAPE over all equipments (e.g., AVG\_CT). %CHWP2 and CWP2 are turned off in the test dataset, and are therefore not reported. %Visualization of SISO models fitting testing data is shown in Figure~\ref{fig:siso}.
%As shown in Figure~\ref{fig:siso}, the relationship between power and VSD speed of CWP follows the affinity laws.
As shown in Table~\ref{tb:mapesiso}, PR in DDO and MLP perform comparably on all modules. PR achieves the smallest average MAPE on CHWP while MLP performs slightly better than PR on CWP and CT. We notice that MLP gets easily trapped in local optimum in training and delivers results with large variance among different models. Considering the difference in model complexity and training complexity, it appears that PR is a better choice for online prediction and optimization. LSTM performs worst on SISO modules. This is due to the lack of temporal dependency in the data. Especially after data cleaning operation that removes the outliers, LSTM has problem to learn the right updating rules over non-consecutive records of the readings. Overall, by applying simple polynomial regression, we are able to achieve small average MAPEs on SISO modules.

\begin{table}
	\small
	\caption{MAPEs(\%) of MISO models}\label{tb:flowtemp}
	\centering
	\scalebox{0.88}{
		\begin{tabular}{|l|c|c|c|}
			\hline
			& MLP (DDO) & PR & LSTM \\ \hline \hline
			CHFM &  \textbf{1.59} & 1.74 & 4.11   \\
			CWFM &  \textbf{1.27} & 1.28 & 2.80   \\
			CWTM &  \textbf{0.92} & 1.33 & 6.46   \\ \hline
			CH1  & 1.82 & 3.26 &    4.22 \\
			CH2  & 2.09 & 2.25 &    2.98 \\
			CH3  & 2.23 & 2.70 &    3.45 \\
			AVG\_CH   & \textbf{2.05}    & 2.73 &    3.55 \\

			%CWFM & \textbf{1.53\%} & 3.85   \\
			%    CHFM & \textbf{1.42\%} & 3.82 \\
			%    CWTM & \textbf{0.38\%} & 6.35 \\
			\hline
	\end{tabular}}
\end{table}

%\begin{table}
%  \centering
%  \begin{tabular}{|l|c|c|c|}
%    \hline
%    % after \\: \hline or \cline{col1-col2} \cline{col3-col4} ...
%      & MLP   & CH3 & Average  \\ \hline \hline
%    MAPE & 0.91\% & 1.59\% & 1.25\%  \\
%    \hline
%  \end{tabular}
%  \caption{MAPE of MISO chiller models for prediction of chiller power}\label{tb:chillerkw}
%\end{table}

\noindent\textbf{Evaluation of MISO Models}
The prediction results of MISO models are summarized in Table~\ref{tb:flowtemp}. For chilled water flow rate, condenser water flow rate, and condenser water temperature prediction, we report the MAPEs of MLP, PR and LSTM. %The corresponding visualization of these MISO models fitting testing data is shown in Figure~\ref{fig:miso}(d-f).
The output of CHFM, CWFM and CWTM are fed into CH models for chiller power prediction. For chillers, we report the MAPE of each chiller and the average MAPE of each model. On all MISO modules, MLP (DDO) performs the best and achieves much smaller MAPEs than those of PR and LSTM. LSTM performs poorly on water flow rate and temperature prediction, which further affects its performance in chiller power prediction. On the contrary, with low MAPEs in water flow rate and temperature prediction, the error propagation of MLP from CHFM, CWFM, CWTM models to the chiller model is minimized.
%For chillers, we report the MAPE of each chiller and the average MAPE in Table~\ref{tb:chillerkw}. CH1 is turned off in test dataset. LSTM is too slow to converge under this setting for chillers. We therefore skip the results of LSTM in this part of the experiments.
%The corresponding visualization of MISO chiller models fitting testing data is shown in Figure~\ref{fig:miso}(a-c).
%Chiller power fluctuates significant along time, showing no clear trend. However,
Our method DDO using MLP is able to capture the dynamics of chiller power quite well, achieving the smallest average MAPE at around 2\%. %This is partly because of the accurate prediction models of condensed and chilled water flow, and condensed water temperature.
%Notice that the prediction error of the third chiller CH3 is higher than CH2 (and CH1 usually). This is because CH3 is mainly a backup chiller with small cooling load, and is operating much less frequently than the other two, which results in fewer samples in the training data.

%\begin{figure*}
%\begin{tabular}{ccc}
%  % after \\: \hline or \cline{col1-col2} \cline{col3-col4} ...
%  \includegraphics[width=2in]{fig/ch1}
%  & \includegraphics[width=2in]{fig/ch2}
%  & \includegraphics[width=2in]{fig/ch3} \\
%  (a) & (b) & (c) \\
%  \includegraphics[width=2in]{fig/cwsfhdr}
%  & \includegraphics[width=2in]{fig/chwsfhdr}
%  & \includegraphics[width=2in]{fig/cwshdr} \\
%  (d) & (e) & (f) \\
%\end{tabular}
%\caption{Visualization of MISO models on the test data}
%\label{fig:miso}
%\end{figure*}

\subsection{Evaluation of Total Power Prediction}

\begin{table}
	\small
	\caption{MAPE of total system power prediction}\label{tb:syskw}
	\centering
	\scalebox{0.88}{
		\begin{tabular}{|l|c|c|c|c|}
			\hline
			& DDO & PR & MLP  & LSTM (black-box)  \\ \hline \hline
			MAPE & 1.86 & 2.24 & \textbf{1.81} & 2.25 \\
			\hline
	\end{tabular}}
\end{table}

Combining all module-wise models, we are able to predict the total power of the chiller plant. The result is reported in Table~\ref{tb:syskw}. Because of the poor performance of LSTM on module-wise models, we separately train a black-box LSTM model to better predict the total system power, directly over all original input variables. The input includes the VSD speeds of pumps and fans, configurations of equipments, weather, system cooling load and the setpoint $chsp$. The MAPE of the black-box LSTM is presented in Table~\ref{tb:syskw}. Among all the models, DDO and MLP perform the best with MLP achieves slightly smaller MAPE. However, the saving in training and optimization time by using PR to model SISO modules in DDO makes up the difference.

%Figure~\ref{fig:total}. The MAPE of total power prediction is $0.89\%$, which indicates minimal error propagation.

%\begin{figure}
%  \centering
%  % Requires \usepackage{graphicx}
%  \includegraphics[width=2in]{fig/totalpower}\\
%  \caption{Prediction of total power of the chiller plant}\label{fig:total}
%\end{figure}

%\subsubsection{Comparison with LSTM}
%Next we compare our approach with a single black-box model for total power prediction. In particular, viewing the chiller plant as a whole and control parameters and weather data as time series, we apply state-of-the-art long short term memory (LSTM) for power prediction. We report the results in ...

\subsection{Evaluation of Real-Time Optimization}

\begin{figure}
	\centering
	\includegraphics[width=.75\columnwidth]{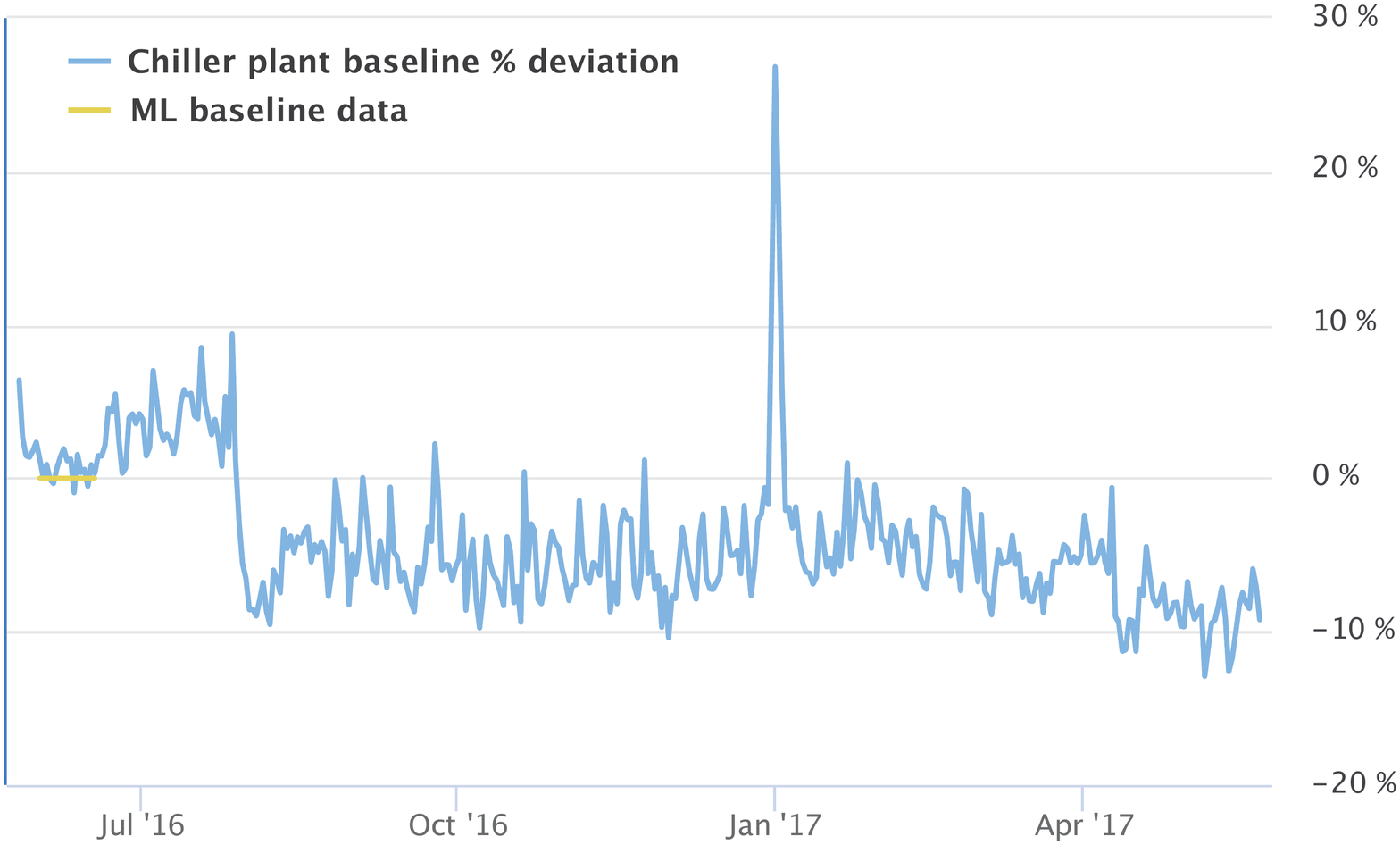}\\
	\vspace{-5pt}
	\caption{Energy saving of real world testing.}\label{fig:saving_result}
	\vspace{-10pt}
\end{figure}

\begin{figure}
	\centering
	\includegraphics[width=.75\columnwidth]{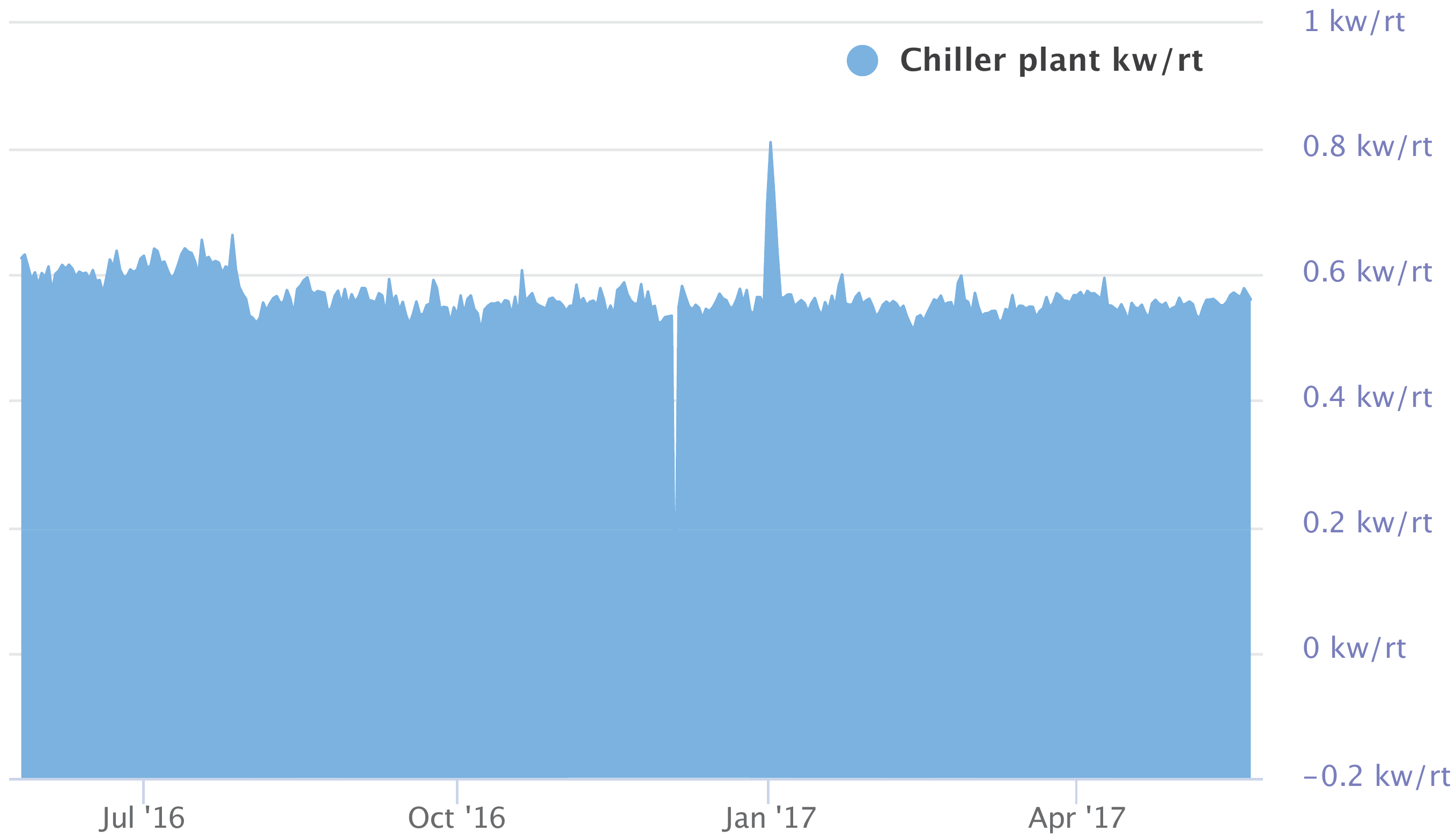}\\
	\vspace{-5pt}
	\caption{Energy efficiency of real world testing.}\label{fig:eff_result}
	\vspace{-10pt}
\end{figure}

We started to apply data-driven real-time optimization on a real world chiller plant in July 2016, and continually improved the optimization technique from August 2016. The optimization results are shown in Figure~\ref{fig:saving_result} and Figure \ref{fig:eff_result}. One of the challenges of performance evaluation is to appropriately normalize the results, in order to fairly compare the energy behavior of different strategies adopted at different time intervals. In our testing, we collect a group of historical data under operation without optimization in June 2016, indicated by \emph{ML baseline data} in Figure \ref{fig:saving_result}. A power consumption prediction model is built over the ML baseline data, using weather and system cooling load as input features. It is then applied on new records to estimate the energy consumption since August (when the chiller plant is controlled by our DDO approach). We then calculate the difference of estimation and actual electricity consumption in percentage and plot them in figure \ref{fig:saving_result}. Obviously, from August 2016, there is a significant drop of energy cost over the estimation on operations without optimization. The average energy saving is between 5\% to 10\%. The energy efficiency is also consistently improved, based on the results in Figure \ref{fig:eff_result}, by using the same estimation approach. The saving is so significant that the chiller plant has saved thousands of dollars by using our DDO approach. Note that there are outlier numbers appearing in January 2017, which are due to equipment replacement.

%The lower part of the figure plots the actual daily total power of the chiller plant from January 2015 to December 2016 in red line. We use data of December 2016 to train a LSTM to predict the total chiller plant power given weather information $rh$, $db$ and $cooling\_load$. It is then applied to predict the daily total chiller plant power of other months from 2015 to 2016. The predicted values are shown as the green line in Figure~\ref{fig:optresult}. Its corresponding MAPEs are shown in the upper part of Figure~\ref{fig:optresult}. As shown in Figure~\ref{fig:optresult}, the predicted total power is lower than that of the actual one, by more than 7\%, from January 2015 to June 2016 when optimization is not applied. This indicates that if similar optimization technique as of Dec 2016 is applied, the total power could be reduced significantly.

\section{Conclusion}\label{sec:conclu}

In this paper, we present our data-driven optimization techniques and report our empirical evaluations of our techniques on real-world chiller plants. Different from existing machine learning approaches, we design the framework and choose the data models based on our domain knowledges. We show that complex machine learning models, such as popular Recurrent Neural Networks, may not be an optimal solution for highly dynamic and complex mechanical systems. Instead, simple models may better capture the actual mechanism within the equipments used in chiller plants. Moreover, active data enrichment is an effective solution to the generalization problem haunting existing approaches with data analysis only. The combination of these new but simple techniques enables our system to accurately capture the running status and dynamically optimize the chiller plant in real time, achieving significant power saving for energy hungry chiller plants.

In the future, we will explore on the following research directions. First, we will attempt to collect more data from sensors in smart buildings, including video feeds from surveillance cameras and audio data from microphones. Such data is helpful to the system to better capture the ongoing activities in the buildings, and finally facilitating better cooling load prediction. Second, we will look into the transfer learning techniques, in order to combine data from multiple chiller plants for more accurate diagnosis analysis. Such methods will be extremely useful, especially for rare fault problems only occurring to each chiller plant a few times in history.

\begin{acks}
This research is funded by the Republic of Singapore's National Research Foundation (NRF) through Building and Construction Authority (BCA)'s Green Buildings Innovation Cluster (GBIC) R\&D Grant, BCA RID 94.17.2.8 (Application No : NRF2015ENC-GBICRD001-065).
\end{acks}

\balance
\bibliographystyle{ACM-Reference-Format}
\bibliography{sigproc}

\end{document}